\pdfoutput=1
\documentclass[twocolumn,secnumarabic,amssymb, nobibnotes, aps, prd,floatfix,superscriptaddress]{revtex4-2} 

\newtheorem{algorithm}{Algorithm}
\usepackage[shortlabels]{enumitem}
\usepackage{enumitem}
\newlist{primenumerate}{enumerate}{1}
\setlist[primenumerate,1]{label={\arabic*$'$}}
\usepackage{caption}
\usepackage{subcaption}
\usepackage{dsfont}
\usepackage{hyperref}
\hypersetup{colorlinks=true, allcolors=blue}

\usepackage{amsmath}
\usepackage{xcolor}
\usepackage{comment}
\usepackage{graphicx}
\usepackage{mathtools}
\usepackage{bbm}
\usepackage{lipsum}

\newcommand{\bs}{{\boldsymbol{\sigma}}}
\newcommand{\bd}{\boldsymbol{\delta}}
\newcommand{\bt}{{\boldsymbol{\theta}}}

\newcommand{\E}{\mathbb{E}}

\newcommand{\Cov}{\mathrm{Cov}}

\newcommand{\la }{\langle}
\newcommand{\ra }{\rangle}
\newcommand{\vertiii}[1]{{\left\vert\kern-0.25ex\left\vert\kern-0.25ex\left\vert #1 
    \right\vert\kern-0.25ex\right\vert\kern-0.25ex\right\vert}}

\DeclareMathOperator*{\argmin}{arg\,min}

\begin{document}

\title{Understanding and eliminating spurious modes in variational Monte Carlo using collective variables}%

\author{Huan Zhang}%
\affiliation{Courant Institute of Mathematical Sciences, New York University, New York 10012, USA}

\author{Robert J. Webber}
\affiliation{Division of Computing and Mathematical Sciences, California Institute of Technology, Pasadena, CA 91125 USA}

\author{Michael Lindsey}%
\affiliation{Department of Mathematics, University of California, Berkeley, CA 94720 USA}

\author{Timothy C. Berkelbach}
\email{tim.berkelbach@gmail.com}
\affiliation{Department of Chemistry, Columbia University, New York, New York 10027, United States}
\affiliation{Center for Computational Quantum Physics, Flatiron Institute, New York, New York 10010, United States}

\author{Jonathan Weare}%
\email{weare@nyu.edu}
\affiliation{Courant Institute of Mathematical Sciences, New York University, New York 10012, USA}

\begin{abstract}
The use of neural network parametrizations to represent the ground state in variational Monte Carlo (VMC) calculations has generated intense interest in recent years.
However, as we demonstrate in the context of the periodic Heisenberg spin chain, this approach can produce unreliable wave function approximations. 
One of the most obvious signs of failure is the occurrence of random, persistent spikes in the energy estimate during training.
These energy spikes are caused by regions of configuration space that are over-represented by the wave function density, which are called ``spurious modes'' in the machine learning literature. 
After exploring these spurious modes in detail, we demonstrate that a collective-variable-based penalization yields a substantially more robust training procedure, preventing the formation of spurious modes and improving the accuracy of energy estimates.
Because the penalization scheme is cheap to implement and is not specific to the particular model studied here, it can be extended to other applications of VMC where a reasonable choice of collective variable is available.
\end{abstract}

\maketitle

\section{Introduction}

Variational Monte Carlo (VMC) is an algorithm for approximating the ground-state energy and wave function of a quantum many-body system~\cite{gubernatis2016quantum,becca2017quantum}. 
As a variational method, VMC seeks the lowest-energy wave function $\psi_\bt(\cdot)$ by minimizing the energy 
with respect to a set of variational parameters $\bt$.
Building on a history of successful VMC applications, researchers have recently introduced neural-network-based families of wave functions that can be evaluated and differentiated efficiently~\cite{carleo2017solving,luo2019backflow,pfau2020ab,hermann2020deep}.
Although neural networks are sufficiently flexible to represent difficult wave functions~\cite{cybenko1989approximation}, 
neural network parameter optimization can be slow or unstable, and parameters can converge to local minima~\cite{yang2020scalable, park2020geometry}, limiting the accuracy that can be practically attained.

\begin{figure}[t!]
    \centering
    \includegraphics[width = 1.0\columnwidth]{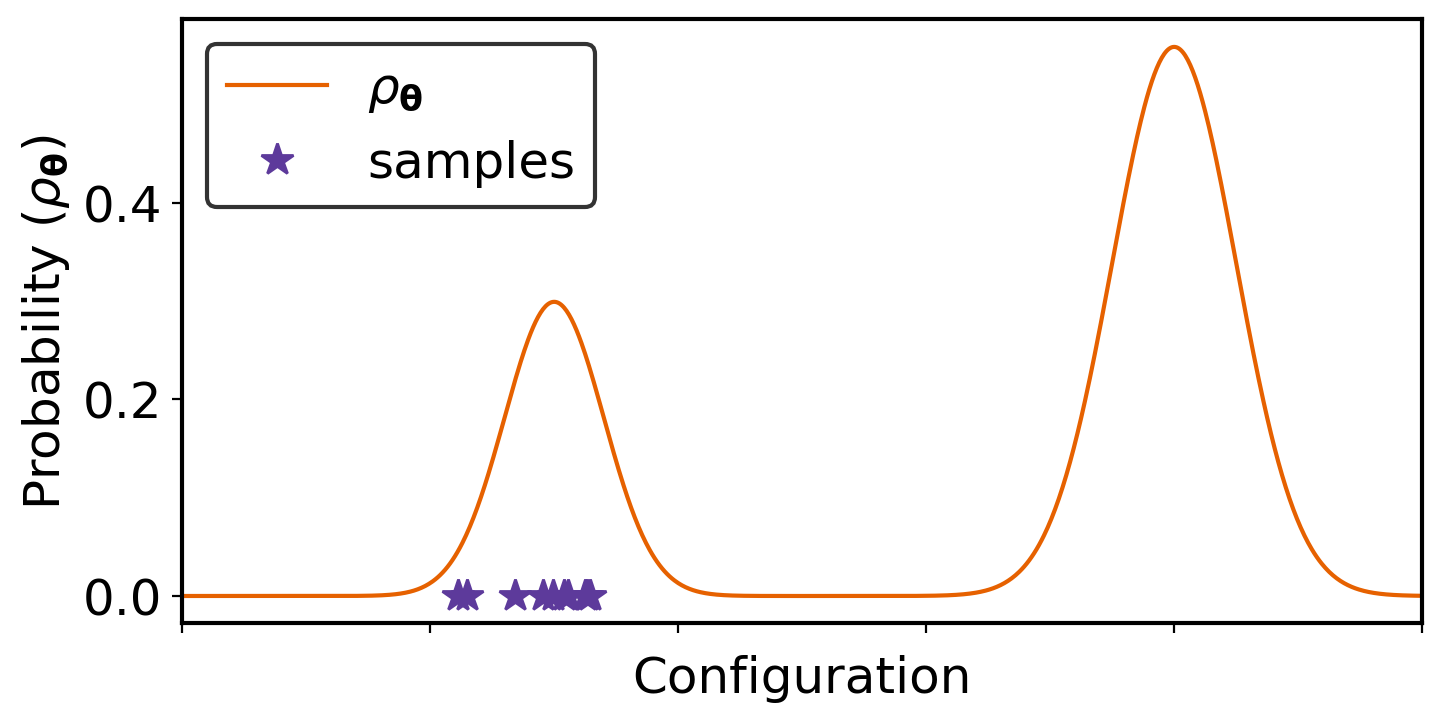}
    \caption{A spurious mode is a high-probability region that is absent in the empirical sample distribution but present in the model distribution.}
    \label{fig:spuriousmode}
\end{figure}

Here, we identify the formation of \textit{spurious modes} as a problem that degrades the accuracy and robustness of neural VMC wave functions.
A spurious mode is defined in the machine learning literature as a high-probability region that is absent in the data distribution but present in the model distribution (see, for example, Ref.~\onlinecite[Sec.~18.1-18.2]{Goodfellow-et-al-2016}).
Like machine learning for probability density estimation,
VMC needs to sample from the wave function probability density $\rho_{\bt} \propto |\psi_\bt(\cdot)|^2$ in order to estimate the energy and energy gradient.
In this work, we show that spurious modes can occur in VMC, i.e., parameter updates yield a wave function probability density that is artificially large in regions far away from the samples, as illustrated schematically in Fig.~\ref{fig:spuriousmode}.
As can be seen in this figure and documented in detail in this article, the formation of spurious modes is only possible because the variational wave function is unconstrained in undersampled regions of configuration space.
As a symptom, the VMC energy estimator typically exhibits a large energy spike when the sampler first encounters a
spurious mode.
The energy spike can persist over thousands of optimization steps, making it difficult to extract a usable energy estimate from VMC.


This work can be viewed as a constructive approach for diagnosing and mitigating generalization error
from limited Monte Carlo sampling, which is already recognized as a challenge for VMC~\cite{westerhout2020generalization}.
Indeed, it is important to address the problem of spurious modes
as neural VMC rapidly grows in popularity and finds applications to systems of ever-increasing complexity~\cite{gubernatis2016quantum,becca2017quantum,carleo2017solving,luo2019backflow,pfau2020ab,hermann2020deep,cybenko1989approximation,yang2020scalable, park2020geometry,Goodfellow-et-al-2016}.
We might hope that enhanced sampling techniques such as parallel tempering and umbrella sampling~\cite{swendsen1986replica,dinner2020stratification,FEbundleprotein1995} could remedy the issue of spurious modes, 
as has been suggested in the machine learning literature~\cite{desjardins2010tempered}.
However, in the context of VMC, we demonstrate that enhanced sampling methods do not solve the problem.

We introduce collective-variable-informed VMC (CV-VMC) as a new effective strategy for addressing spurious modes in VMC.
To our knowledge we are the first to link the spurious modes in VMC to the appearance of energy spikes and also the first to offer a satisfactory resolution.
The main idea of our CV-VMC framework is to exploit a well-chosen one-dimensional collective variable (CV) that distinguishes physically reasonable configurations from physically unreasonable configurations. 
More specifically, we generate a pool of samples across a range of CV values, and we use the samples to penalize large wave function densities in physically unreasonable regions of configuration space.

CV-VMC is simple and cheap to implement, and it leverages physical intuition to accelerate and improve VMC optimization,
inspired by the use of CVs in guiding free energy calculations~\cite{hydrophobicMC1979,Berneche2001CVUS}. 
While the design of appropriate CVs must be calibrated based on the type of problem, this work advances a general and modular framework for incorporating \emph{a priori} intuitions about the wave function into the optimization procedure itself.

The paper is structured as follows.
Section \ref{sec:overfitting} presents the model problem of interest as well, as well as the traditional VMC optimization approach. In Section~\ref{sec:results}, we study the appearance of energy spikes in VMC, connect the spikes to the existence of spurious modes, and show the inadequacy of generic enhanced sampling techniques for addressing this problem.
Section \ref{sec:solution} introduces and tests the CV-VMC approach for eradicating the spurious modes.
Section \ref{sec:conclusion} concludes.

%

\section{Preliminaries}\label{sec:overfitting}



Throughout this work, we apply VMC to the antiferromagnetic Heisenberg model for $N$ spin-$1\slash 2$ particles in a one-dimensional periodic chain, defined
by the Hamiltonian
\begin{align}
\label{eq:hamiltonian}
    \hat{H} = \sum_{i=1}^{N} \bigl(
    \hat{\sigma}_i^x \hat{\sigma}_{i+1}^x
    + \hat{\sigma}_i^y \hat{\sigma}_{i+1}^y
    + \hat{\sigma}_i^z \hat{\sigma}_{i+1}^z \bigr),
\end{align}
where $\hat{\sigma}_i^x$, $\hat{\sigma}_i^y$, and $\hat{\sigma}_i^z$ are the Pauli operators for the $i$th spin. Here and throughout, periodic boundary conditions are implied via the identification $\hat{\sigma}_{N+1}=\hat{\sigma}_{1}$. 
For our variational wave function, we use the neural quantum state ansatz~\cite{carleo2017solving} inspired by the restricted Boltzmann machine (RBM)~\cite{tieleman2008training}. This ansatz is a two-layer (or one-hidden-layer) neural network that has been widely used in VMC in recent years~\cite{webber2021rayleigh}. 
We refer to our ansatz as the RBM throughout. 
Working in the many-body basis of spin configurations $\bs$ defined as simultaneous eigenstates of the operators $\{\hat{\sigma}_i^z\}$ with eigenvalues $\{\sigma_i\}$ (dropping the $z$ indicator for notational simplicity), an RBM wave function can be written as
\begin{equation}
\psi_{\bt}(\bs) =
    \sum_{\{h_k\}} \exp\Bigl(\sum_{i=1}^{N} a_i \sigma_i
    +
    \sum_{k=1}^{M} b_k h_k
    +
    \sum_{ik} W_{ki}h_k \sigma_i\Bigr)
\end{equation}
where $\sigma_i \in \{-1,+1\}$ are the spin variables,
$h_k \in \{-1,+1\}$ are an additional set of $M$ hidden spin variables,
and $\bt=\{\boldsymbol{a},\boldsymbol{b},\boldsymbol{W}\}$ are the variational parameters. 
Summing over the hidden spins $h_k$ and
enforcing the translational symmetry that is exhibited by the
exact ground state gives the modified RBM ansatz,
\begin{align}
\psi_{\bt}(\bs) =
    \prod_{k=1}^M \prod_{j=1}^N \cosh \Bigl(b_k + \sum_{i=1}^N W_{ki}\sigma_{i+j} \Bigr),
\end{align}
which reduces the set of variational parameters to
$\bt=\{\boldsymbol{b},\boldsymbol{W}\}$, and, again, periodic boundary conditions
are implied.

For fixed parameters $\bt$, the energy can be calculated as
\begin{equation}
\label{eq:energy}
E = \frac{\langle \psi_\bt, \hat{H}\psi_\bt\rangle}
    {\langle \psi_\bt, \psi_\bt\rangle}
    = \sum_{\{\sigma_i\}} E_\mathrm{loc}(\bs) \rho_\bt(\bs)
\end{equation}
where $E_\mathrm{loc}(\bs) = (\hat{H}\psi_\bt)(\bs)/\psi_\bt(\bs)$ is the local energy,
$\rho_\bt(\bs) \propto |\psi_\bt(\bs)|^2$ is the normalized
probability density, and we have adopted the inner product notation
\begin{equation}
\langle \psi , \phi \rangle = \sum_{\{\sigma_i\}} \overline{\psi(\bs)} \phi(\bs).
\end{equation}

It remains to optimize the parameters $\bt$ in the RBM ansatz to minimize the energy functional $E$. 
Here, we use the stochastic reconfiguration (SR)~\cite{becca2017quantum} algorithm.
In the SR method, the parameter update $\bd$ can be derived by minimizing a cost function
\begin{align}
    \label{eq:SRloss}
    \frac{\la \psi_{\bt+\bd}, H \psi_{\bt+\bd}\ra }{ \la \psi_{\bt+\bd}, \psi_{\bt+\bd}\ra}
    -
    \frac{1}{\epsilon}
    \left(
    \frac{\left|\langle\psi_\bt, \psi_{\bt+\bd}\rangle\right|}
    {\left\Vert \psi_\bt\right\Vert \left\Vert \psi_{\bt+\bd}\right\Vert}
    \right)^{2},
\end{align}
which contains the usual energy expression~(\ref{eq:energy}) and an additional penalization term that prevents large wave function updates.
After differentiating the cost function~\eqref{eq:SRloss} and performing algebraic manipulations (see \cite{webber2021rayleigh}), this approach leads to the following algorithmic approach to VMC.

\begin{algorithm}[VMC via SR]
Choose the parameter update $\bd$ to solve
\begin{equation}
    (\boldsymbol{S}+\eta \boldsymbol{I})\bd=-\epsilon \boldsymbol{g}.
\end{equation}
Here, $\eta \geq 0$ is a nonnegative parameter chosen to make $\boldsymbol{S}+\eta \boldsymbol{I}$ positive definite.
The energy $E$, gradient vector $\boldsymbol{g}$, and overlap matrix $\boldsymbol{S}$ are defined by
\begin{subequations} \label{eq:VMC}
\begin{align}
    E &= \mathbb{E}_{\rho_\bt} \left[E_\mathrm{loc}(\bs)\right], \\
    g_i&=\Cov_{\rho_\bt}\left[ \frac{\partial_{\theta_i} \psi _{\bt}(\bs)}{\psi _{\bt}(\bs)} ,\frac{H\psi _{\bt}(\bs)}{\psi _{\bt}(\bs)}\right], \\
    S_{ij}&=\Cov_{\rho_\bt}\left[ \frac{\partial_{\theta_i} \psi_{\bt}(\bs)}{\psi _{\bt}(\bs)} ,\frac{\partial_{\theta_j} \psi _{\bt}(\bs)}{\psi _{\bt}(\bs)} \right],
\end{align}
\end{subequations}
where $\mathbb{E}_{\rho_\bt}$ and $\Cov_{\rho_\bt}$ indicate the expectation value and covariance with respect to the current probability distribution $\rho_\bt(\bs)$.
\label{alg:VMC}
\end{algorithm}   

Because of the high dimensionality, the averages appearing above are estimated stochastically. With SR, this leads to the following iterative VMC strategy:
\begin{enumerate}
    \item Draw samples from the probability distribution
    \begin{equation}
    \rho_{\bt} (\bs)\propto |\psi _{\bt}(\bs)|^2
    \end{equation}
    \item Use the samples to provide an energy estimate $\hat{E}$, gradient estimate $\hat{\boldsymbol{g}}$, and overlap estimate $\hat{\boldsymbol{S}}$ by Eqs.~\eqref{eq:VMC}.
    \item Update $\bt$ by solving the regularized linear system
    \begin{equation}
    \label{eq:linear}
    (\hat{\boldsymbol{S}}+\eta \boldsymbol{I})\bd=-\epsilon \hat{\boldsymbol{g}}
    \end{equation}
    and setting $\bt \leftarrow \bt+\bd$.
\end{enumerate}

In this paper, we generate samples from the wave function density $\rho_{\bt} (\bs)\propto |\psi _{\bt}(\bs)|^2$ using a Metropolis-type Markov chain Monte Carlo (MCMC) sampler.
Because the Hamiltonian~\eqref{eq:hamiltonian} conserves the total magnetization, we focus on the sector with $\sum_i \sigma_i = 0$. Our sampler starts from a uniformly distributed configuration within the subspace and attempts to swap a randomly chosen $+1$ spin with a randomly chosen $-1$ spin.
During each iteration (defined as one parameter update), we generate data by running MCMC for 2000 Metropolis steps,
and we subsample the data once every 100 steps to reduce the storage and computation costs. 
We typically run $n = 100$ independent MCMC walkers of this type and pool together the resulting data to calculate parameter updates.
However, as enhanced sampling alternatives, we also experiment with the parallel tempering \cite{swendsen1986replica} and umbrella sampling \cite{dinner2020stratification} methods described in Section \ref{sec:enhanced}.


Throughout this work, we use $M=5N$ hidden spins.
We initialize our neural network wave function parameters as independent complex-valued $\mathcal{N}\left(0, 0.001\right)$ random variables.
We increase the penalization parameter $\epsilon$ from $\epsilon = 0.001$ to $\epsilon = 0.01$ at a geometric rate over first $500$ iterations, after which it is held constant, and we use $\eta=0.001$ for all iterations. When a large parameter update occurs, we restore $\epsilon$ to its initial value and restart the geometric progression.
More discussion about the initialization and parameter choices for stochastic reconfiguration for models of this type can be found in Ref.~\onlinecite{webber2021rayleigh}.

\section{A study of spurious mode formation} \label{sec:results}

\subsection{Energy spikes and spurious modes}
\label{sec:spurious}

\begin{figure}[t!]
    \centering
    \includegraphics[width = 1.0\columnwidth]{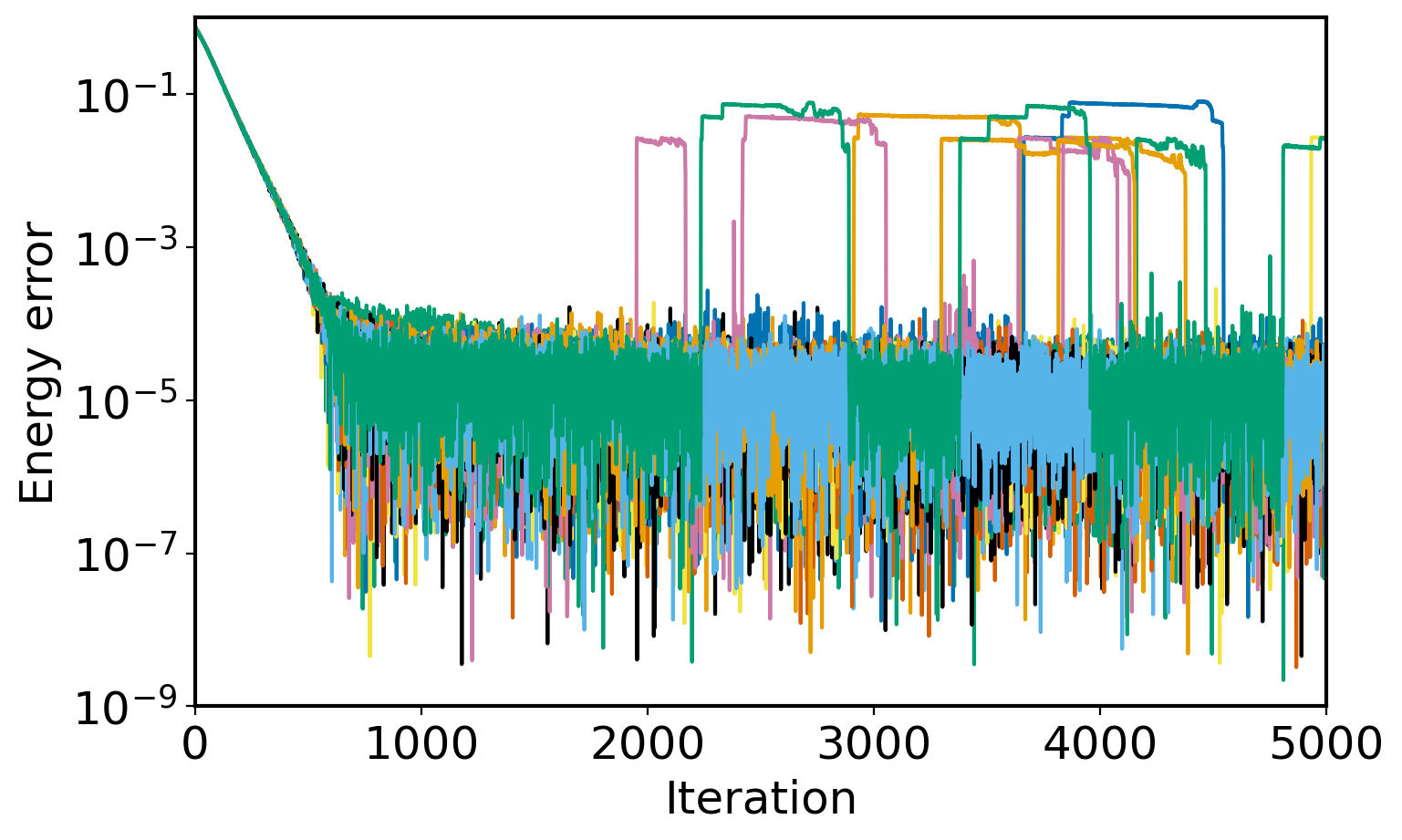}
    \caption{Per-site error in VMC energy estimates obtained over $20$ independent runs
    for a chain of $N=100$ spins.
    The exact energy is computed using the Bethe ansatz \cite{bethe1931theorie}.}
    \label{fig:spikes}
\end{figure}

When we optimize our VMC wave function for 5000 iterations, we obtain the results depicted in Figure \ref{fig:spikes}.
The VMC energy error decreases over the initial 1800 iterations and begins to exhibit high-frequency fluctuations on the scale of $10^{-4}$.
Then, during iterations 1800--5000, large sustained energy spikes occur in 9 out of 20 independent VMC training runs.
In several training runs, the energy spikes appear more than once.

The occurrence of the energy spikes is concerning for two reasons.
First, the repeated spikes make it difficult to decide when the VMC statistics converge and when to stop training.
Second, spikes may arise unexpectedly when an apparently converged VMC wave function is used for downstream investigations, including the computation of other observables besides the energy~\cite{misawa2019mvmc} and the refinement of the energy estimate by diffusion Monte Carlo~\cite{needs2009continuum}.
We explore this possibility in more detail in Sec.~\ref{sec:importance}.

The frequency of the spikes depends on the number $N$ of sites in the chain and the amount of sampling performed at each training iteration. In Table~\ref{tab:test} we report the number of energy spikes observed in 20 training runs for various chain lengths $N$ and numbers $n$ of parallel MCMC walkers.
When $N$ is small ($N \leq 50$), 
there are fewer spikes.
When the number of walkers is small ($n \leq 20$), the energy spikes rarely if ever occur,
but in this case the VMC energy estimates are inaccurate, with variance 10 times higher than in the $n = 100$ case.
If infinitely many MCMC steps were performed between parameter updates and $\epsilon$ were sufficiently small, energy spikes could not occur~\cite{webber2021rayleigh}. In practice, however, VMC is carried out far from this limit.
With $n = 500$ parallel walkers, we still observe many training runs with energy spikes.

\begin{table}[b!]
    \centering
    \begin{tabular}{l|c|c|c}
        ~ & $N=50$ & $N=100$ & $N=200$ \\
        \hline
        $n=20$~ & $0$ & $0$ &  $0$ \\
        $n=50$~ & $1$ & $8$ & $6$ \\
        $n=100$~ & $0$  & $9$ & $9$ \\
        $n=200$~ & $1$  & $8$ & $14$ \\
        $n=500$~ & $0$ & $5$ &  $12$
    \end{tabular}
    \caption{The number of training runs (out of $20$) exhibiting energy spikes.
    Energy spikes are identified by per-site energy estimates with error $\geq 10^{-3}$.}
    \label{tab:test}
\end{table}

\begin{figure}[t!]
    \centering
    \includegraphics[width = 1.0\columnwidth]{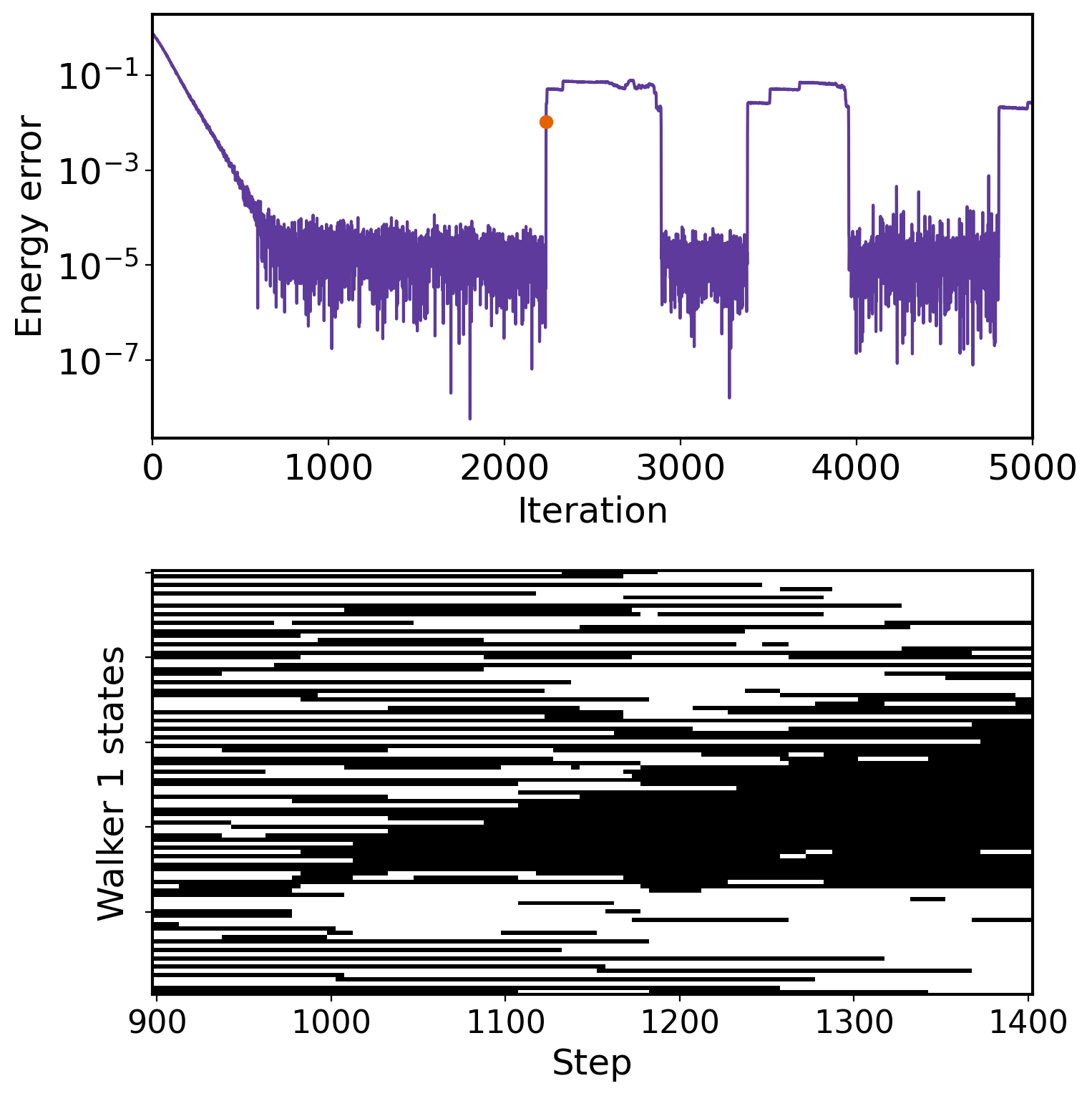}
    \caption{Top: per-site energy error, with an orange dot marking iteration 2236 in which the first energy spike occurs.
    Bottom: states of walker 1 during steps 900--1400 of iteration 2236, with black indicating up spins and white indicating down spins.}
    \label{fig:states}
\end{figure}

We can zoom in on an energy spike to understand the phenomenon better.
Fig.~\ref{fig:states} presents a typical VMC training run exhibiting energy spikes. 
The first spike occurs at iteration 2236, which is marked by the orange dot in the upper panel.
The lower panel shows the cause of the energy spike.
During the sampling stage of iteration 2236, a single MCMC chain (`walker 1') transitions suddenly from an antiferromagnetic state (characteristic of the ground state of the antiferromagnetic Heisenberg model studied here) to a ferromagnetic state with two domain walls.

\begin{figure}[t!]
    \centering
    \includegraphics[width = 1.0\columnwidth]{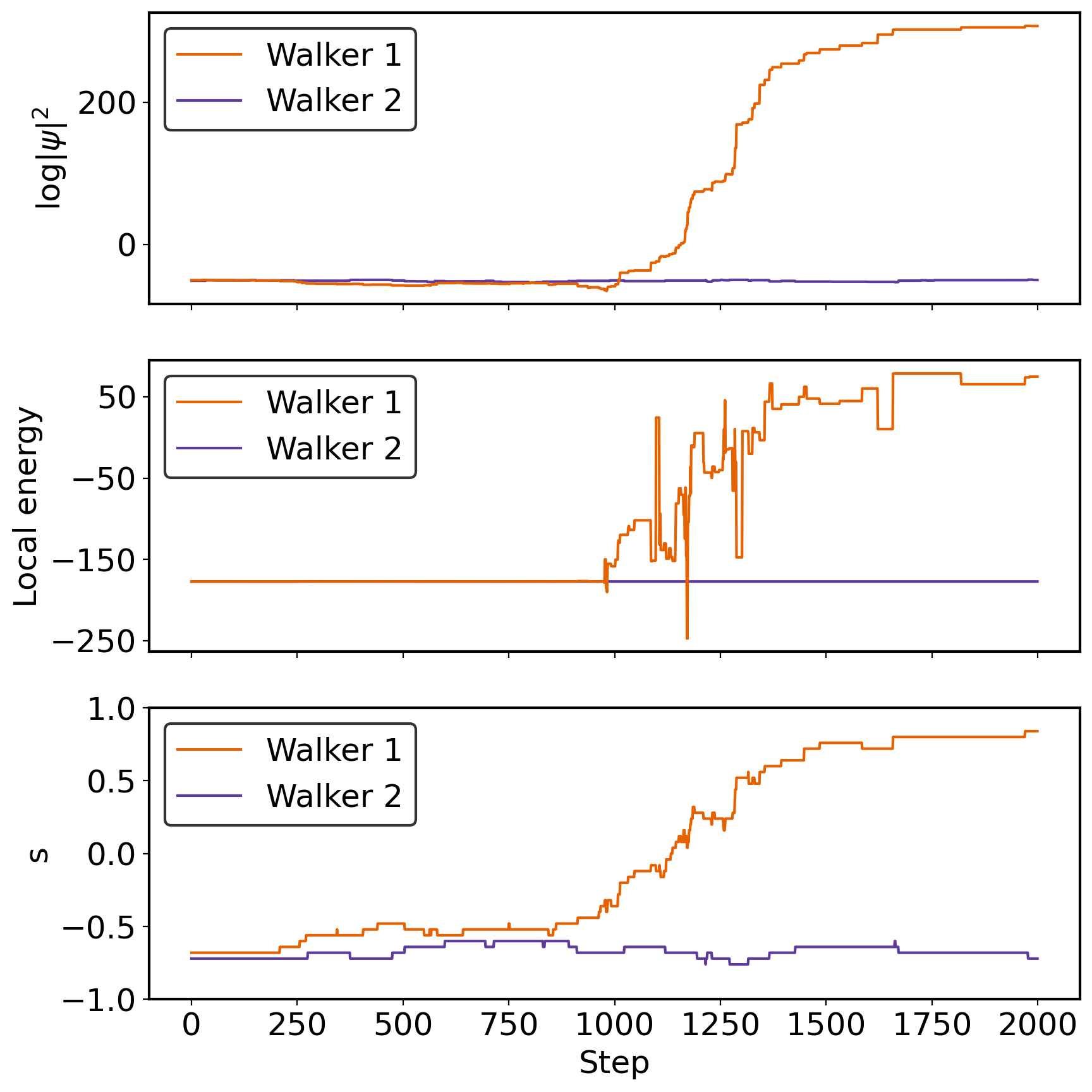}
    \caption{Values of $\log |\psi|^2$ (top), local energy (middle), and $s$ (bottom) for walker 1 and walker 2.}
    \label{fig:walker}
\end{figure}

The statistics of walker 1 are presented in more detail in Fig.~\ref{fig:walker}.
We notice that walker 1 experiences an abrupt increase in its estimated probability density $\rho_{\bt}$ (consistent with a spurious mode), 
as well as a large increase in its local energy, yielding the spike in the energy estimate.
Motivated by the trajectory in Fig.~\ref{fig:states}, we characterize this transition with a CV that captures the local magnetic ordering,
\begin{align}
    s(\boldsymbol{\sigma})\equiv \frac{1}{N} \sum_{i=1}^{N} \sigma_i \sigma_{i+1}.
    \label{eq:reactioncoordinate}
\end{align}
The collective variable $s$ ranges from $-1$ (antiferromagnetic), to $0$ (nonmagnetic), to $+1$ (ferromagnetic).
For the antiferromagnetic Heisenberg Hamiltonian~\eqref{eq:hamiltonian}, we expect the ground state to be predominantly supported by configurations with $s<0$.
We see in Fig.~\ref{fig:walker} that the value of $s$ for walker 1 sharply increases exactly when the energy spike occurs.

For comparison, we also plot data for one of the MCMC chains (`walker 2') that shows no abrupt change in either the wave function magnitude or local energy.
Most of the $n = 100$ MCMC walkers have profiles similar to walker 2.
Apart from walker 1, only two other MCMC chains ever enter the $s>0$ region and contribute to the energy spike, starting at iterations 2241 and 2332.

\begin{figure*}[ht!]
    \centering
    \includegraphics[width = 1.0\textwidth]{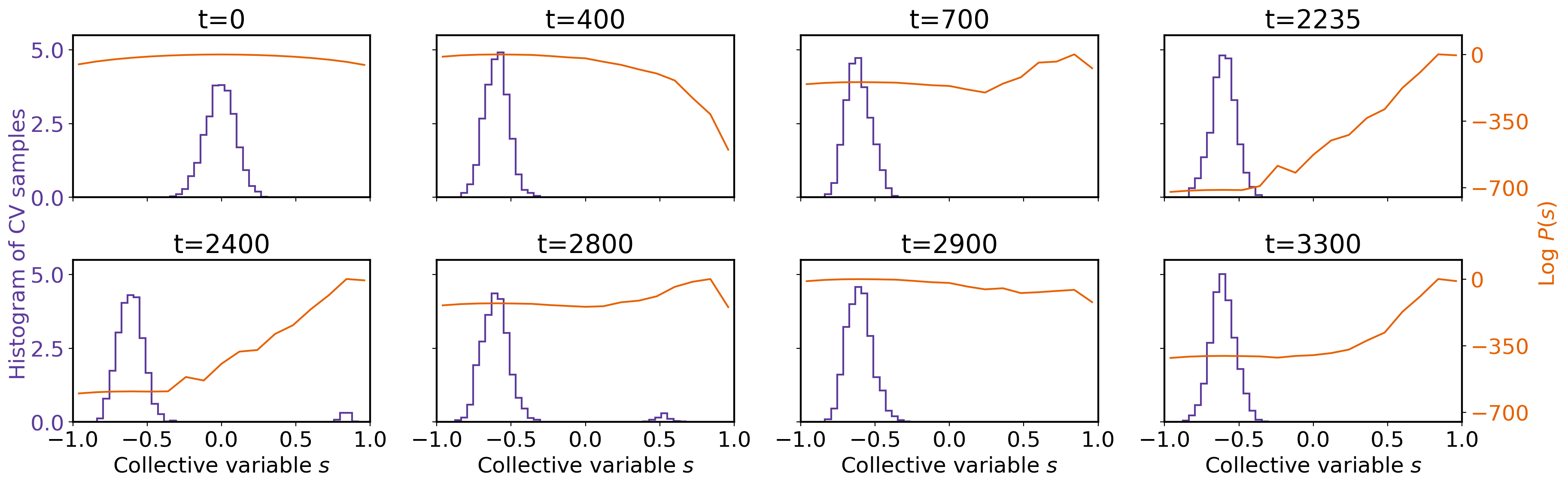}
    \caption{Empirical histogram of CV samples (purple) and the marginal probability density $P(s)$ of the CV (orange). %
    }
    \label{fig:change_hist}
\end{figure*}

We loosely define a \textit{spurious mode} as a collection of configurations $\boldsymbol{\sigma}$ for which
the wave function probability density $\rho_{\bt}(\bs)$ is large and $s(\bs) > 0$, the latter of which implies that the local
energy will be large for this Hamiltonian. 
As seen in Fig.~\ref{fig:walker}, the energy spike begins when walker 1 suddenly encounters a configuration in a spurious mode.

Fig.~\ref{fig:change_hist} charts the emergence of a spurious mode over thousands of optimization steps. The orange line indicates the marginal probability density for $s$, 
\begin{equation}
P(s') = \sum_{\boldsymbol{\sigma}} 
    \delta_{s',s(\bs)} \, \rho_{\bt}(\bs)
\end{equation}
where $\delta_{s_1, s_2}$ indicates the Kronecker delta.
$P(s)$ is estimated using an enhanced   sampling procedure described in Appendix~\ref{app:EMUS}. 
At time $t = 0$, the MCMC walkers are randomly initialized with a symmetric distribution of $s$ values.
However, by time $t = 400$, all the walkers have moved into the $s < 0$ region that is the physically relevant mode of the antiferromagneic Heisenberg model.
Meanwhile, the marginal probability density $P(s)$ in the undersampled $s > 0$ region starts to increase and forms a spurious mode.
Starting at iteration $t = 2236$, several of the walkers find their way across the energy barrier to the spurious mode, and they become stuck due the high wave function probability there.
The local energies in the $s > 0$ region are far higher than the ground state energy, leading to a dramatic energy spike.
The bottom row of Fig.~\ref{fig:change_hist} demonstrates the spurious mode disappearing via sampling, i.e., walkers return towards $s<0$ and the marginal probability density $P(s)$ is corrected.
In the final panel, a spurious mode has reappeared and will lead to another energy spike.


\subsection{Robustness testing identifies spurious modes}
\label{sec:importance}

With a view toward downstream tasks,
we assess the reliability of the optimized wave function (i.e., the possibility of a spurious mode) at a given VMC iteration using the following robustness test:
\begin{enumerate}
    \item[i.] With the parameters $\bt$ held fixed to their value at the current iteration, we choose 100 configurations from the final states of the MCMC chains used to train the optimized wave function.  Starting from these states, we generate an additional $2 \times 10^6$ Metropolis steps, saving samples every $10^5$ steps.
    \item[ii.] We compute an energy estimate using the resulting $2\times 10^3$ sample configurations.
    \item[iii.] We repeat steps (i) and (ii) 10 times and compare the resulting 10 energy estimates.
    
\end{enumerate}

\begin{figure}[t]
    \centering
    \includegraphics[width = 1.0\columnwidth]{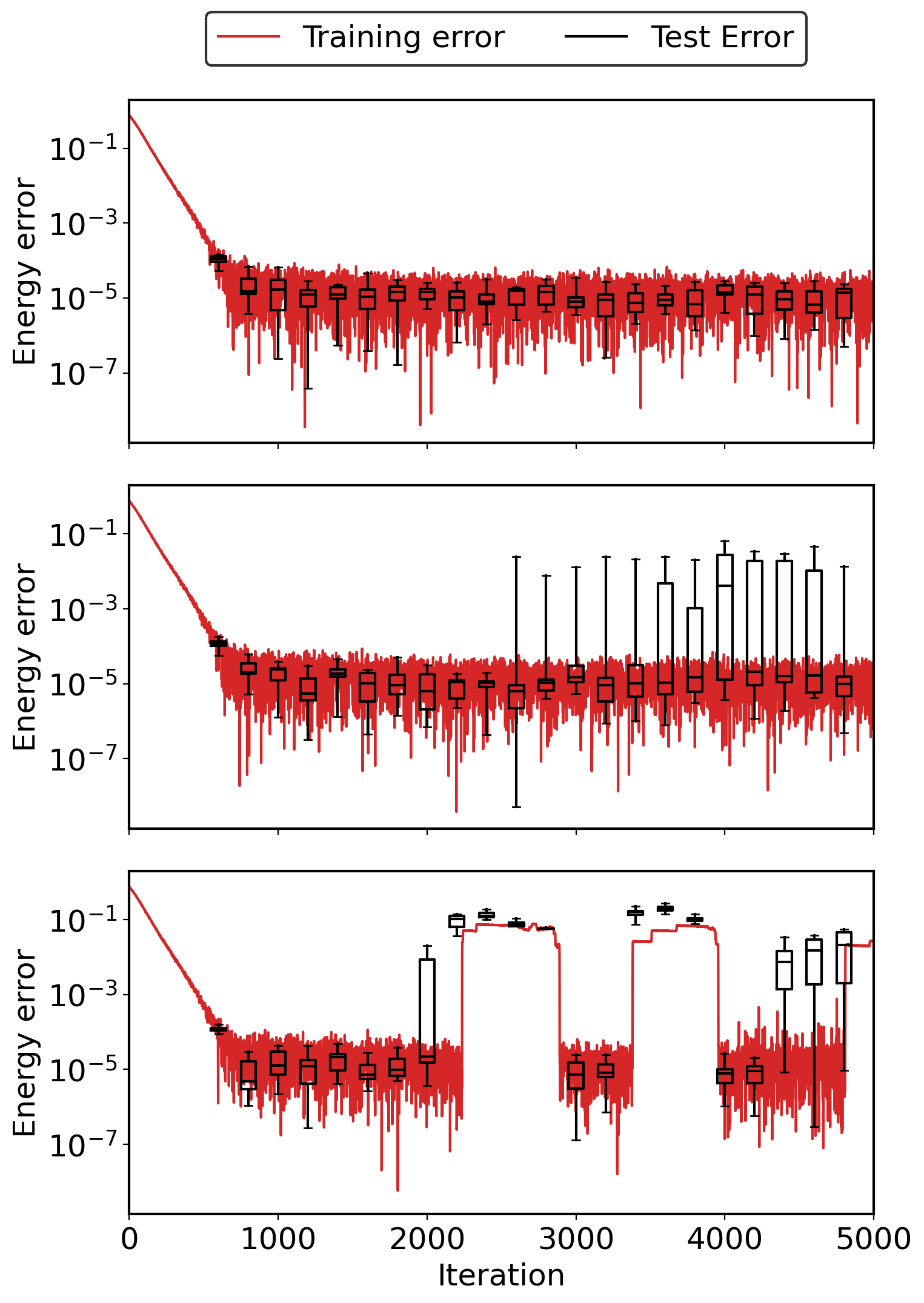}
    \caption{Per-site energy errors during training (red) and during robustness testing (box-and-whisker plots, black) for three independent VMC training runs.}
    \label{fig:robust}
\end{figure}

The mean and variance calculated from steps (ii) and (iii) are indicative of the global quality of the current wave function and the existence of spurious modes.
Figure \ref{fig:robust} presents results from three independent VMC optimizations, showing the instantaneous energy error (red) and the statistics of the energy errors calculated as described above (box-and-whisker plots, black).

There are several possible outcomes:
the VMC training shows no spikes and the robustness testing shows no spikes (top panel),
the VMC training shows no spikes but the robustness testing shows spikes (middle panel),
or both training and robustness testing show spikes (bottom panel).
We infer that the optimized wave function in the top
panel has no spurious modes, the optimized wave function in the middle panel
develops a spurious mode around iteration 2500 that is \textit{unobserved}
during training even after 5000 iterations, and the optimized wave function in the third panel develops
multiple spurious modes that are also evident in training.

These results highlight the fact that accurate energies during training do not guarantee accurate energies during testing.
Indeed, we tested the robustness of the 20 VMC wave functions obtained via the 20 independent VMC runs shown in Fig.~\ref{fig:spikes}.
Although 11 of the 20 VMC runs did not show any energy spikes during training, 
5 of these 11 VMC runs exhibited energy spikes in testing, with an energy spike here defined as a per-site energy estimate with error $\geq 10^{-3}$.

\subsection{Enhanced sampling does not prevent spurious mode formation}
\label{sec:enhanced}

We have demonstrated that undersampled regions are prone to the formation of spurious modes in the VMC wave function. Enhanced sampling methods, 
which have been developed specifically to increase sampling in low-probability regions without sacrificing the in-principle exactness of Monte Carlo estimation, seem to offer a straightforward remedy. We explore two of the most commonly used enhanced sampling methods, parallel tempering and umbrella sampling, but find that in fact they do not prevent the formation of spurious modes.

Parallel tempering is a popular enhanced sampling method across a wide range of applications~\cite{Earl2005partemp}, including in RBM training~\cite{desjardins2010tempered}. 
The previous paper~\cite{webber2021rayleigh} applied parallel tempering 
to improve sampling in VMC.  In parallel tempering, multiple MCMC chains indexed by $k$ are simulated, each sampling from a density proportional to $\rho_{\bt}^{\beta_k}$, with the values of the $\beta_k$ spaced to cover the interval $[0,1]$.  The states of these chains are periodically exchanged, ideally allowing the chain sampling $\rho_{\bt}$ itself to escape local maxima of the density~\cite{Earl2005partemp}. Only data from the chain sampling $\rho_{\bt}$ is used to compute averages. 

In our tests, parallel tempering causes the spikes to occur more frequently but subside more rapidly, as shown in Figure \ref{fig:PTenergy}. (See Appendix~\ref{app:PT} for more details.)
Once the spurious mode is fully formed, parallel tempering can transport the chain there promptly, and subsequent parameter updates can partially correct the wave function within the spurious mode. But before the spurious mode is fully formed, samples from $\rho_{\bt}$ are concentrated in the $s<0$ region, and the emerging spurious mode has no effect on training.
In short, parallel tempering does nothing to prevent the emergence of the spurious modes in the first place.
\begin{figure}[t]
    \centering
    \includegraphics[width = 1.0\columnwidth]{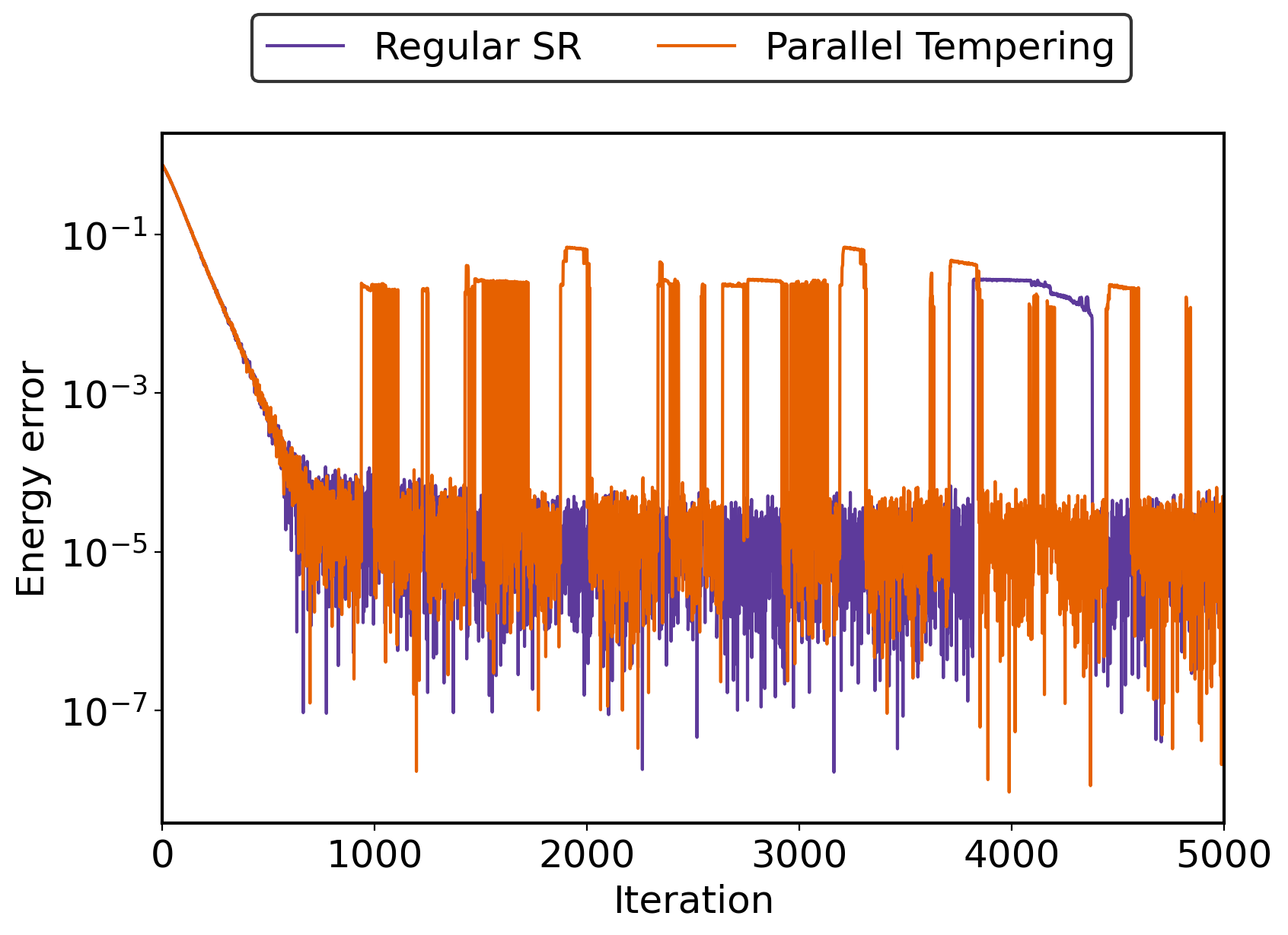}
    \caption{VMC with parallel tempering (orange) leads to shorter, more frequent spikes than VMC with direct MCMC sampling (purple).}
    \label{fig:PTenergy}
\end{figure}

We also consider the umbrella sampling method, which has been used in free energy calcuations for decades~\cite{Kastner2011US,thiede2016eigenvector}. Umbrellas sampling in the form used here is a stratified sampling method for general averages, as suggested and explored in Ref.~\onlinecite{dinner2020stratification}. In umbrella sampling, each MCMC sampler is restricted to remain within a range, or `window,' of possible $s$ values. By covering the range of all possible $s$ values with such windows (16 windows in our tests), sampling resolution is increased in the $s>0$ region relative to unbiased MCMC sampling. Statistical weights are then assigned to samples to correct for the biased sampling distribution.  See Appendix~\ref{app:EMUS} or Ref.~\onlinecite{dinner2020stratification} for further details of the method. 

\begin{figure}[h!]
    \centering
    \includegraphics[width = 1.0\columnwidth]{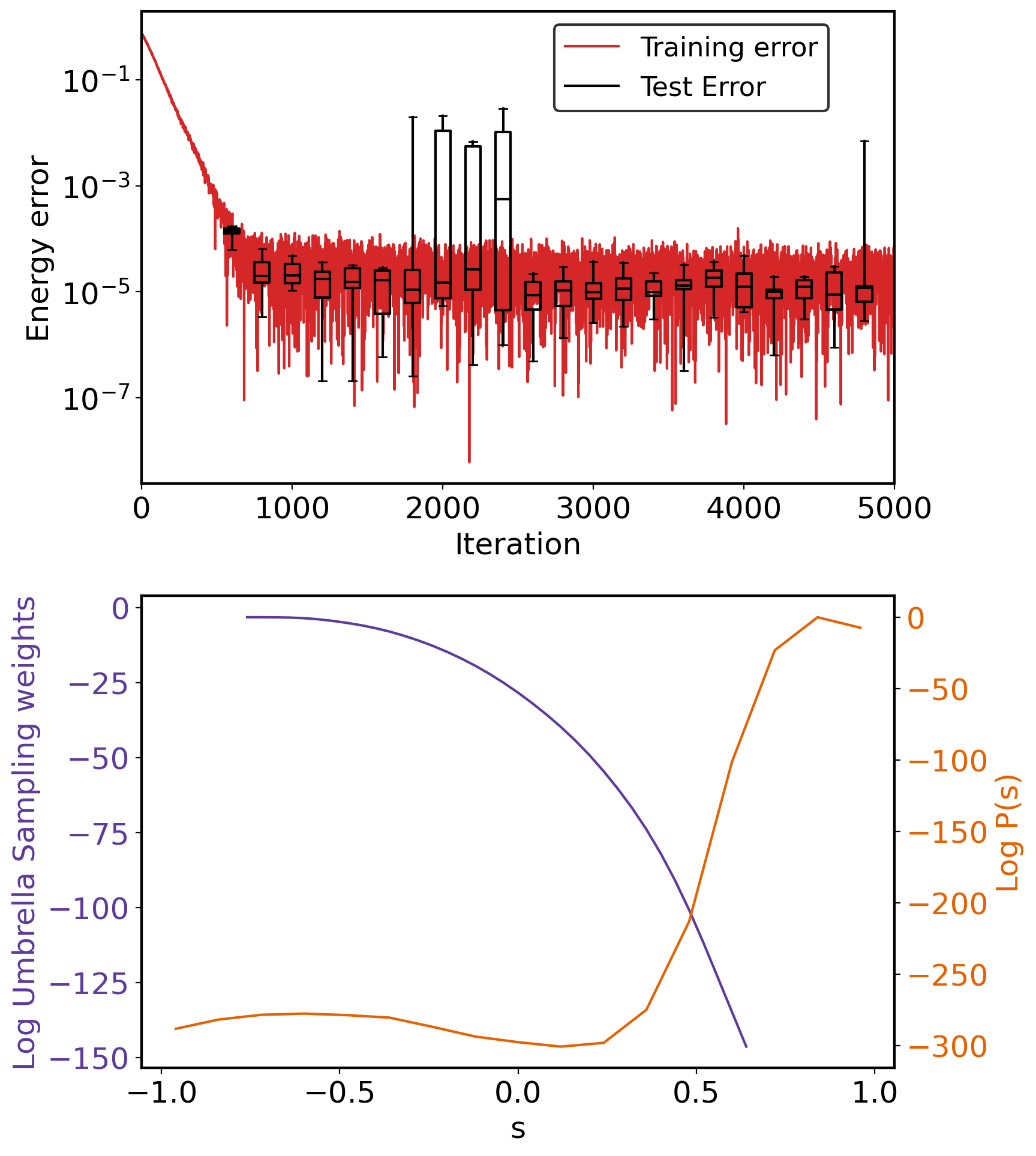}
    \caption{Top: per-site energy errors during the training (orange) and during the testing (box-and-whisker plots, black). Bottom: weights assigned to each of the restrained umbrella sampling MCMC samplers (purple), along with the marginal probability density $P(s)$ of the CV (orange).
   }
    \label{fig:CVUS_2}
\end{figure}

While umbrella sampling does eliminate visible energy spikes during training, we find that, like parallel tempering, it does not prevent the formation of spurious modes. A typical training run and the results of our robustness test 
are shown  Fig.~\ref{fig:CVUS_2} (top panel).
Because the proposal distribution used in our Metropolis sampling scheme favors moves toward $s=0$, the MCMC samplers restrained to regions of higher $s$ suffer from very low acceptance rates (0.01--0.05) and tend to discover the spurious mode well after it has formed. Having not discovered the region of artificially high probability, umbrella sampling assigns very small statistical weight to samples in the higher $s$ region, and the emerging spurious mode has no impact on parameter updates. In Fig.~\ref{fig:CVUS_2} (bottom panel), for a representative choice of parameters, we validate this claim by comparing the average statistical weight assigned to samples at each value of $s$  to the marginal density $P(s)$ of $\rho_\bt$.
Details of this calculation can be found in Appendix~\ref{app:EMUS}.

Robustness tests for the parallel tempering trained wave function approximation are carried out exactly as described in Section~\ref{sec:importance}. Robustness tests for the umbrella sampling-trained wave function approximation require an additional resampling of the final states of the MCMC chains used in training because of the statistical weights assigned to samples in umbrella sampling.
For more details see Appendix~\ref{app:robust}.

\section{Collective-variable-informed VMC}
\label{sec:solution}


In this section we introduce a new term into the objective function of VMC meant to penalize the formation of spurious modes. The new term, which we call the `spurious mode functional,' eliminates spurious modes by confining the wave function to the physically reasonable region of $s$ values. Moreover it is quick to evaluate and differentiate.
In the next subsection, we mathematically motivate our method, which we call collective-variable-informed VMC (CV-VMC). The new objective that defines CV-VMC is optimized via SR. Results from numerical experiments are presented in the following subsection.

\subsection{Mathematical motivation} \label{sec:cv-vmc}
Define the spurious mode functional $L_\text{CV}$ by 
\begin{align}
    L_\text{CV}(\bt)=-\Delta\Cov_{\bs\sim p}\left[\log  |\psi_\bt(\bs)|^2, \mathbbm{1}_{s(\bs)< c} \right]
\label{eq:LCV}
\end{align}
where we have used the symbol $\mathbbm{1}_I$ to indicate the function that is 1 when $\boldsymbol{\sigma}\in I$ and $0$ otherwise.
This expression involves a reference density $p$, which is fixed in advance. Unlike the wave function density $|\psi_\bt|^2$, the reference density $p$ is not adjusted during training. Samples can therefore be drawn from $p$ once, prior to training.  Effective choices of $p$ will place non-negligible weight in the $s>0$ region where $|\psi_\bt|^2$ should be (but is not always) very small. We describe the particular choice of $p$ used in our experiments in detail in Appendix \ref{app:CV-VMCpool}.

The spurious mode functional \eqref{eq:LCV} can be viewed as rewarding the concentration of wave function mass in the physical region $ \{\bs : s(\bs) < c\}$, or equivalently as penalizing the accumulation of mass outside of this region.
The CV-VMC objective function is the sum of the ordinary energy functional and the new spurious mode functional, and the optimization problem becomes
\begin{align}
        \argmin _\bt     \left\{\frac{\la \psi_{\bt}, H \psi_{\bt}\ra }{ \la \psi_{\bt}, \psi_{\bt}\ra}+L_\text{CV}(\bt)\right\}.
        \label{eq:cv_vmc}
\end{align}

We optimize this new objective via SR in which the parameter update $\bd$ is derived by minimizing the cost function
\begin{align}
    \label{eq:CVloss}
    \frac{\la \psi_{\bt+\bd}, H \psi_{\bt+\bd}\ra }{ \la \psi_{\bt+\bd}, \psi_{\bt+\bd}\ra}+L_\text{CV}(\bt+\bd)
    -
    \frac{1}{\epsilon}
    \left(
    \frac{\left|\langle\psi_\bt, \psi_{\bt+\bd}\rangle\right|}
    {\left\Vert \psi_\bt\right\Vert \left\Vert \psi_{\bt+\bd}\right\Vert}
    \right)^{2}.
\end{align}
The Wirtinger derivative~\cite{Wirtinger1927} of $L_\text{CV}(\bt)$ is
\begin{align}
\label{eq:cv_update_penalty}
    \frac{\partial}{\partial \overline{\bt}}L_\text{CV}(\bt)=-\Delta  \Cov_{\bs \sim p}\left[ \frac{\partial_{\bt_i} \psi _{\bt}(\bs)}{\psi _{\bt}(\bs)} , \mathds{1}_{s(\bs) < c}\right],
\end{align}
which yields the following algorithm.

\begin{algorithm}[CV-VMC via SR]
    Choose the parameter update $\bd$ to solve
    \begin{equation}
        (\boldsymbol{S}+\eta \boldsymbol{I})\bd
        = \epsilon (\Delta \tilde{\boldsymbol{g}} - \boldsymbol{g}).
    \end{equation}
    Here, $\eta \geq 0$ is a nonnegative parameter chosen to make $\boldsymbol{S}+\eta \boldsymbol{I}$ positive definite.
    The gradients $\tilde{\boldsymbol{g}}$ and $\boldsymbol{g}$
    and overlap matrix $\boldsymbol{S}$ are defined by
\begin{align}
    \begin{split}
        \tilde{\boldsymbol{g}}_i &= \Cov_{\bs \sim p}\left[ \frac{\partial_{\bt_i} \psi _{\bt}(\bs)}{\psi _{\bt}(\bs)} , \mathds{1}_{s(\bs) < c}\right], \\        
        \boldsymbol{g}_i &=\Cov_{\bs \sim |\psi|^2}\left[ \frac{\partial_{\bt_i} \psi _{\bt}(\bs)}{\psi _{\bt}(\bs)} ,\frac{H\psi _{\bt}(\bs)}{\psi _{\bt}(\bs)}\right], \\
        \boldsymbol{S}_{ij} &=\Cov_{\bs \sim |\psi|^2}\left[ \frac{\partial_{\bt_i} \psi _{\bt}(\bs)}{\psi _{\bt}(\bs)} ,\frac{\partial_{\bt_j} \psi _{\bt}(\bs)}{\psi _{\bt}(\bs)} \right].
    \end{split}
\end{align}
\end{algorithm}
Relative to the SR implementation of ordinary VMC, the only difference is the replacement of the gradient $\boldsymbol{g} \leftarrow \boldsymbol{g}-\Delta\tilde{\boldsymbol{g}}$.



\begin{figure}[h!]
    \centering
    \includegraphics[width = 1.0\columnwidth]{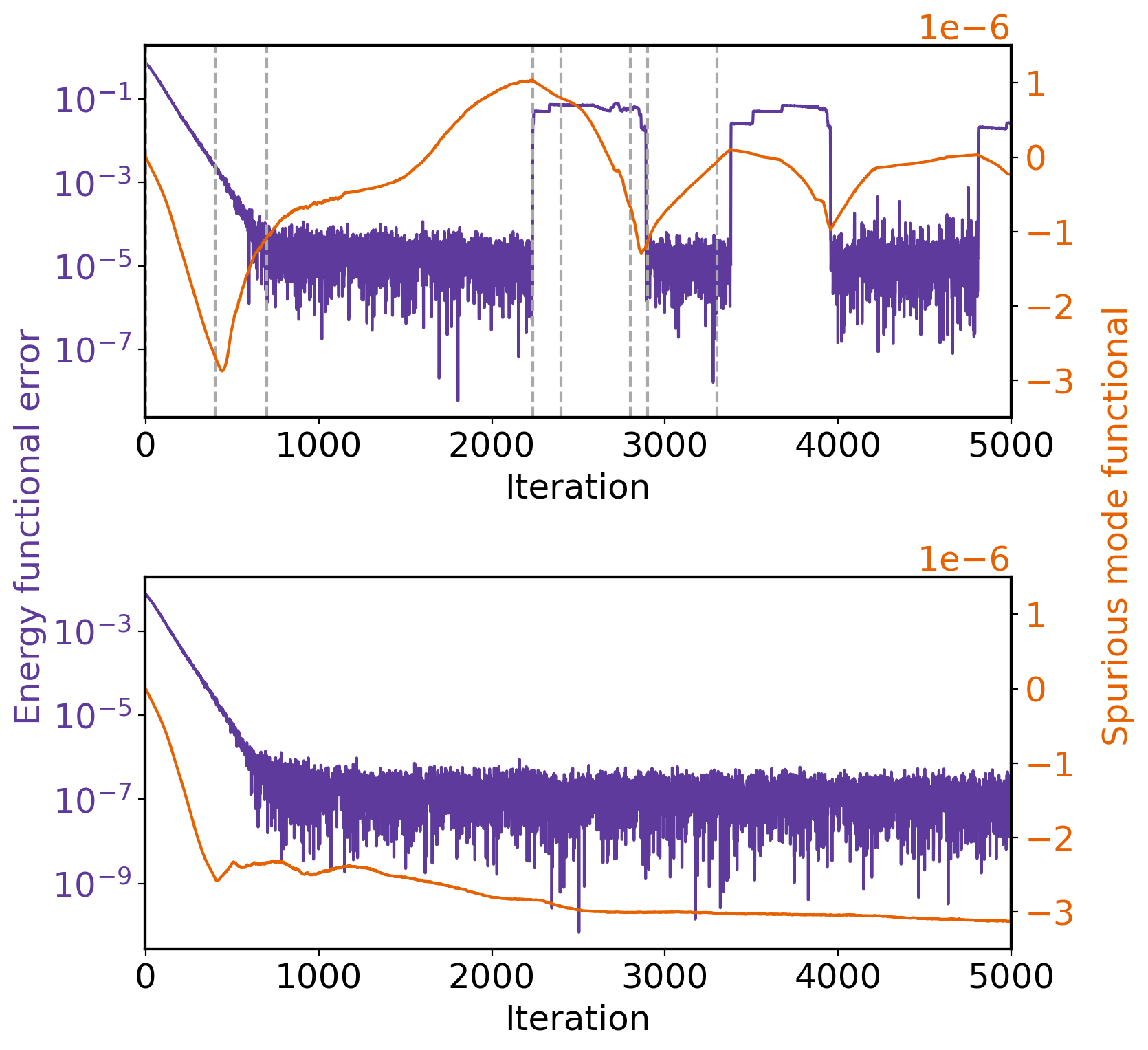}
    \caption{Top: per-site energy functional error (purple) and per-site spurious mode functional (orange) during the same VMC training run depicted in Fig.~\ref{fig:change_hist}. The iterations associated with panels in Fig.~\ref{fig:change_hist} are indicated by vertical dashed lines. Bottom: same quantities during CV-VMC training with the same random seed.}
    \label{fig:CV-VMC-functional}
\end{figure}

To validate the success of the spurious mode functional in detecting spurious modes, in the top panel of Fig.~\ref{fig:CV-VMC-functional} we plot both the energy functional and the spurious mode functional over the course of the \emph{same} VMC (\emph{not} CV-VMC) training run depicted in Fig.~\ref{fig:change_hist}. Evidently the spurious mode functional tracks the formation of a spurious mode even before the energy spikes appear in the energy estimate. In the bottom panel of Fig.~\ref{fig:CV-VMC-functional}, we plot the same quantities over the course of a CV-VMC run, noting that CV-VMC controls the spurious mode functional throughout the training. 

\subsection{Numerical results} \label{sec:numerical test}
 
\begin{figure}[ht!]
    \centering
    \includegraphics[width = 1.0\columnwidth]{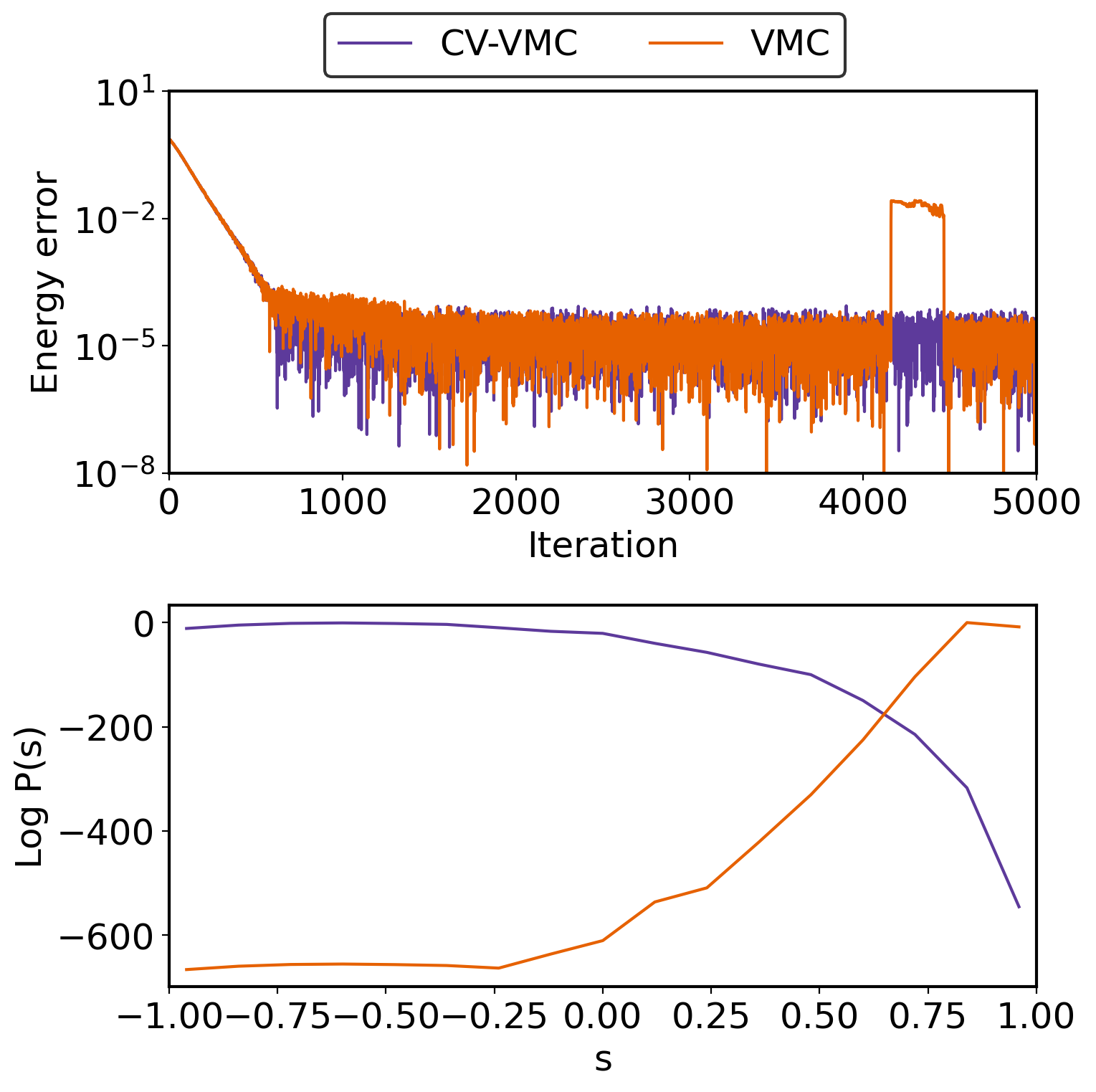}
    \caption{Top: per-site energy errors for VMC and CV-VMC.
    Bottom: marginal density $P(s)$ of CV for the VMC and CV-VMC wave functions at iteration 3000.}
    \label{fig:CV-VMC_VMC}
\end{figure}

Until commented otherwise, the results of this section concern Heisenberg spin chains with $N=100$ spins, and estimates are computed with $n=100$ independent walkers. Fig.~\ref{fig:CV-VMC_VMC} presents results comparing VMC and CV-VMC.
The energy errors are nearly the same for the two optimization approaches except that VMC exhibits a large spike during iterations 4100--4500 (top panel).
Additionally, even before VMC exhibits an energy spike,
there is a spurious mode in the VMC wave function (bottom panel).
In contrast, CV-VMC does not lead to any energy spikes, and the CV-VMC wave function is physically reasonable, quickly losing probability mass
as $s$ increases.

To produce the data in Fig.~\ref{fig:CV-VMC_VMC},
we have chosen the reference density $p$ to be stratified across a wide range of $s$ values (see Appendix \ref{app:CV-VMCpool}) so
that the spurious modes can be readily identified from data.
To speed up the computations,
we have prepared a large pool of $5 \times 10^5$ samples before beginning the VMC optimization and then subsampled 2000 randomly chosen configurations to determine each parameter update.
The resulting algorithm negligibly increases the computational cost relative to VMC.

Although in principle VMC with perfect sampling converges to a local minimum of the energy, CV-VMC should converge to a different fixed point due to the additional gradient term $\Delta \tilde{\boldsymbol{g}}$. To evaluate this new source of error, we show energy errors for different choices of $\Delta$ and $c$ in Table~\ref{tab:energy_delta_c}.
For comparison, the average energy error over 11 cherry-picked standard VMC runs that happened to avoid energy spikes is $6.7\times 10^{-6}$.

\begin{table}[htbp!]
\parbox{.45\linewidth}{
\centering
    \begin{tabular}{c|c}
        $\Delta$ & Energy error \\
        \hline
        $2.5\times 10^{-6}$~ & $7.5 \times 10^{-6}$  \\
        $5.0\times 10^{-6}$~ & $7.9 \times 10^{-6}$  \\
        $1.0\times 10^{-5}$~ & $1.2 \times 10^{-5}$   \\
        $2.0\times 10^{-5}$~ & $3.8 \times 10^{-5}$  
    \end{tabular}
    }
\parbox{.45\linewidth}{
\centering    
    \begin{tabular}{c|c}
        $c$ & Energy error \\
        \hline
        $-0.2$~ & $8.1 \times 10^{-6}$  \\
        $0.0$~ & $7.9 \times 10^{-6}$  \\
        $0.2$~ & $1.0 \times 10^{-5}$   \\
        $0.4$~ & $8.8 \times 10^{-6}$  
    \end{tabular}
    }
    \caption{Mean per-site energy error of 20 independent runs for different $\Delta$ and $c$. For each run, the energy is estimated by taking the average of sampled local energies of all walkers in iterations 1000-5000. In the left table, $c \equiv 0$. In the right table, $\Delta\equiv5.0\times 10^{-6}$.} 
    \label{tab:energy_delta_c}
\end{table}

\begin{figure}[htbp!]
    \centering
    \includegraphics[width = 1.0\columnwidth]{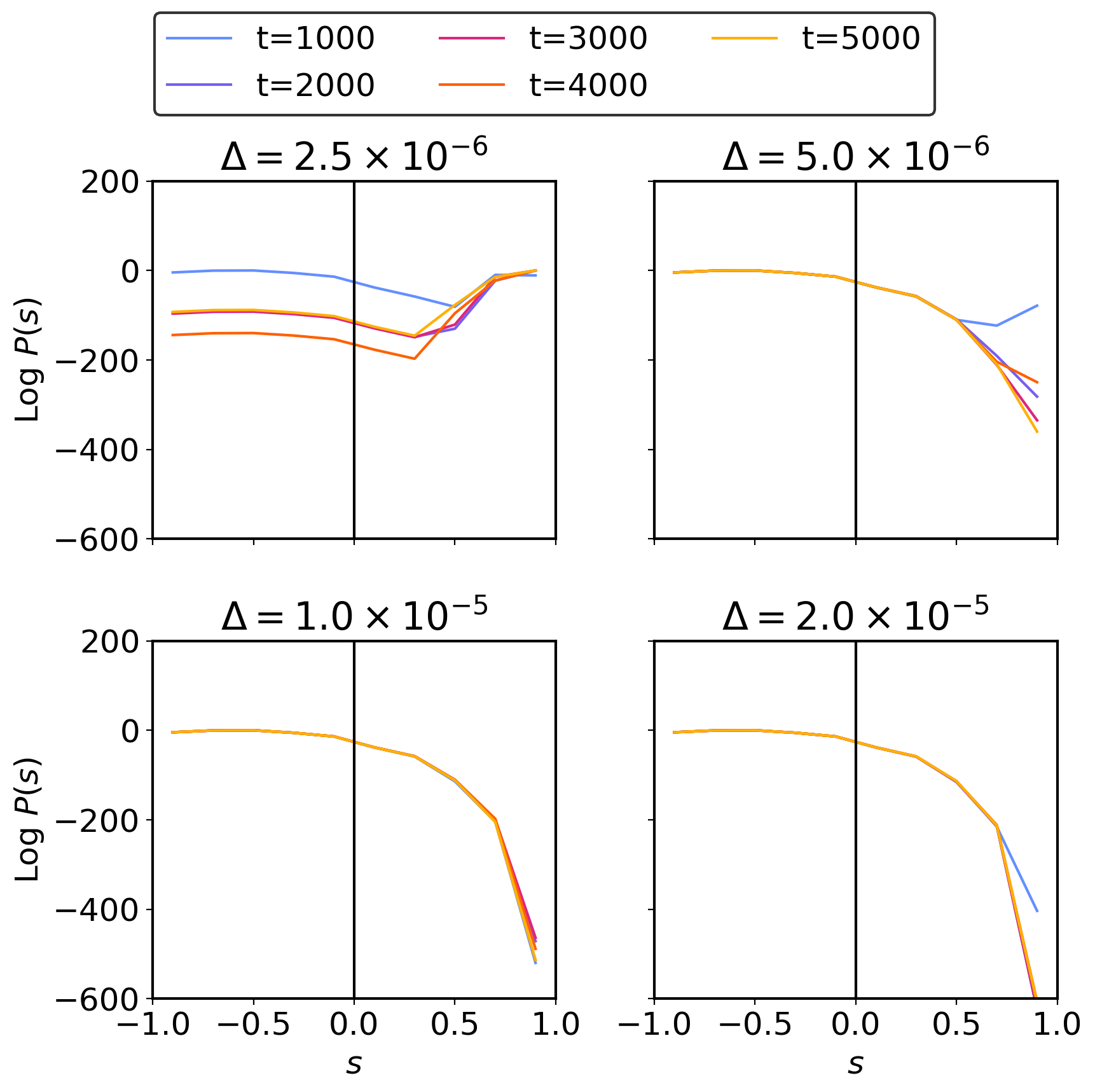}
    \caption{The CV-VMC scheme with $c = 0$ and increasing penalization strengths $\Delta$.
    The colored lines shows the marginal density $P(s)$ at different times during the training. 
    }
    \label{fig:CV-VMC_epsilon}
\end{figure}

First we discuss the impact of varying $\Delta$ for fixed $c=0$. 
All of the runs in the table \emph{happened to avoid} spikes, though we find that for the smallest value $\Delta = 2.5 \times 10^{-6}$, an unexplored spurious mode does develop. Indeed, when $\Delta$ is small (e.g., $\Delta = 2.5 \times 10^{-6}$), the energy estimate is close to that of the reference VMC energy, but the penalty is too weak to eliminate the spurious mode, as shown in Fig.~\ref{fig:CV-VMC_epsilon}. As $\Delta $ gradually increases and the penalization becomes stronger, the energy estimate error increases (see Table~\ref{tab:energy_delta_c})  and the spurious mode is eliminated.  In our tests, we find that $\Delta = 5 \times 10^{-6}$ is large enough to reliably eliminate spurious modes but still small enough to preserve the quality of the energy estimate. In general we recommend choosing $\Delta$ sufficiently large to eliminate energy spikes in both training and testing, but no larger.

Next we discuss the impact of varying $c$ for fixed $\Delta = 5 \times 10^{-6}$. First we rewrite $\Delta\tilde{\boldsymbol{g}}$ as
\begin{multline} \label{eq:penalty_g}
      \Delta\tilde{\boldsymbol{g}}
      = \Delta \E_{\bs\sim p}\bigl[\mathbbm{1}_{s(\bs) < c}\bigr]
      \ \times \\
      \bigl( \E_{\bs\sim p}\bigl[ \overline{\partial_{\bt}\log \psi_\bt} \,\big|\, s(\bs)< c \bigr]-\E_{\bs\sim p}\bigl[\overline{\partial_{\bt}\log \psi_\bt}\bigr] \bigr).
\end{multline}
Let us consider the behavior of this expression in the limits $c \to \pm 1$. When $c\to -1$, $\E_{\bs\sim p}[\mathbbm{1}_{s(\bs) < c}]\to 0$ because the CV cannot take values less than $-1$. Meanwhile, when $c\to 1$, we have \begin{equation}
    \E_{\bs\sim p}\bigl[ \overline{\partial_{\bt}\log \psi_\bt} \,\big|\, s(\bs)< c \bigr]-\E_{\bs\sim p}\bigl[\overline{\partial_{\bt}\log \psi_\bt}\bigr]
    \rightarrow 0
\end{equation}  
because the CV cannot take values greater than $1$. In both cases, Eq.~\eqref{eq:penalty_g} implies that $\tilde{\boldsymbol{g}} \to 0$.

\begin{figure}[h!]
    \centering
    \includegraphics[width = 1.0\columnwidth]{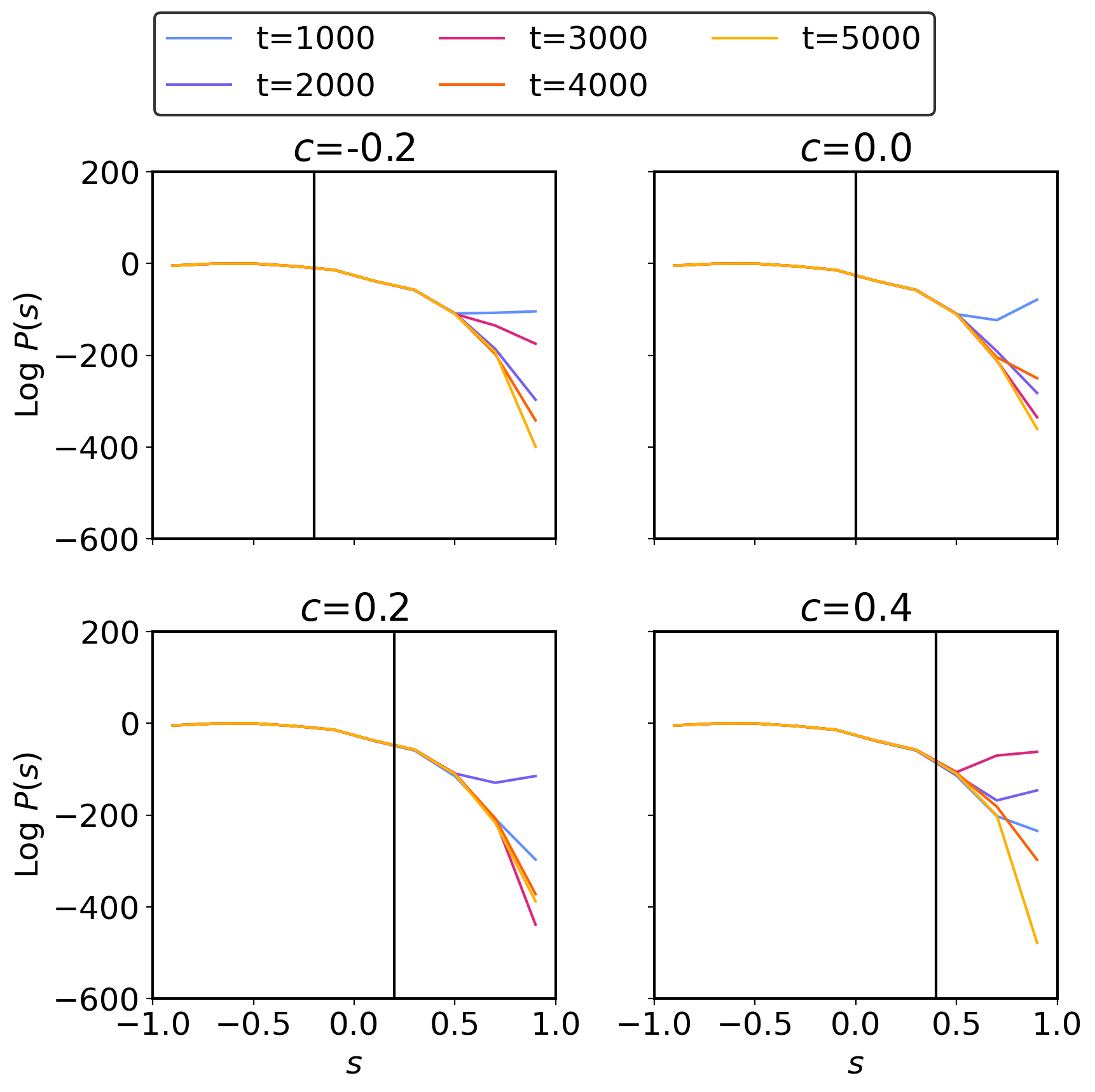}
    \caption{CV-VMC results with $\Delta = 5 \times 10^{-6}$ and different cutoff values $c$ as indicated by black vertical lines. 
    }
    \label{fig:CV-VMC_c}
\end{figure}

In our numerical experiments, we tested the values $c=-0.8, -0.6,...,0.8$, fixing $\Delta=5\times 10^{-6}$ throughout. For values $c\leq -0.4$ and $c\geq0.8$, we observe the formation of spurious modes. 
When $c=-0.6$, for example, we find that one in 20 independent runs develops a spurious mode and exhibits energy spikes.
In contrast, when $-0.2 \leq c \leq 0.4$, we plot the marginal density $P(s)$ of the CV in Fig.~\ref{fig:CV-VMC_c},
and there is no evidence of a spurious mode developing.
Meanwhile, Table~\ref{tab:energy_delta_c} confirms that the energy error of CV-VMC is robust to the choice of $c$ for $c=-0.2,0.0,0.2,0.4$, i.e., in the range where spurious modes are avoided. Ideally, one can set $c$ as a threshold beyond which the true ground state assigns a negligible amount of probability mass.

Finally we demonstrate that CV-VMC outperforms VMC if we do not allow for cherry-picking of spike-free runs. This conclusion holds even when we consider a larger system, consisting of $N=200$ spins. In Table~\ref{tab:d200}, we show energy estimate errors for both methods in the cases $N=100,200$, averaged over several independent runs. Energy spikes contaminate the results for standard VMC, yielding results with per-site energy error on the order $ \sim 10^{-3}$.

\begin{table}[htbp!]
    \centering
    \begin{tabular}{l|c|c|}
        ~ & $N=100$ & $N=200$ \\
        \hline
        VMC~ & $4.970\times 10^{-3}$ & $3.282\times 10^{-3}$  \\
        CV-VMC~ &$7.950\times 10^{-6}$ & $1.390\times 10^{-5}$ 
    \end{tabular}
    \caption{VMC and CV-VMC per-site energy errors. Each number is obtained by averaged over 20 independent runs.}
    \label{tab:d200}
\end{table}

\section{Conclusion}\label{sec:conclusion}

Variational Monte Carlo is a major framework for calculating ground state wave functions of many-body systems, 
but it is limited by the potentially long autocorrelation time of Markov chain Monte Carlo sampling.
Incomplete sampling of the configuration space can result in incorrect gradient estimation, local overfitting, and large, random energy spikes.
These issues cannot always be addressed by simply increasing the length of the Markov chains used for estimation~\cite{yang2020scalable, park2020geometry,webber2021rayleigh}, or even by applying enhanced sampling approaches.

Modifying the objective function is a more promising strategy for improving VMC robustness.
Here, we have proposed collective-variable-informed VMC (CV-VMC) as a specific approach for modifying the objective function to incoporate \emph{a priori} intuitions about the wave function.
We have demonstrated the advantage over standard VMC for the periodic Heisenberg spin chain with $100$--$200$ spins in a series of numerical experiments.
The CV-VMC approach has a negligible added cost relative to standard VMC and is broadly applicable to ground-state estimation problem in which a reasonable choice of collective variable is available.


\section{Acknowledgement}
This work was supported in part through the NYU IT High Performance Computing resources, services, and staff expertise. 
H.Z. and J.W. acknowledge support  from 
the Advanced Scientific Computing
Research Program within the DOE Office of Science through award DE-SC0020427. H.Z. was also supported by  the National Science Foundation through award  DMS-1913129. 
R.J.W. was supported by the Office of Naval Research through BRC award N00014-18-1-2363 and the National Science Foundation through FRG award 1952777, under the aegis of Joel A. Tropp. 
M.L. acknowledges the support of the National Science Foundation under Award No. 1903031 as well as his host institution for this fellowship, the Courant Institute of Mathematical Sciences, New York University.
J.W. was also supported by the National Science Foundation
through award DMS-2054306.

\begin{appendix}



\section{Parallel tempering} \label{app:PT}
In parallel tempering (or replica exchange) one generates samples using several MCMC simulations, each targeting a distribution of the form $\pi_i\propto |\psi_\bt|^{2/T_i}$. The constants $T_i$ are commonly referred to as temperatures and the chain targeting $T_i=1$ samples from $\rho_\theta$. Periodically an exchange of states is proposed between samplers targeting neighboring temperatures and accepted or rejected according to the Metropolis criterion~\cite{Earl2005partemp}.
In Section~\ref{sec:enhanced}, we use 9 temperatures 1, 1.4, 2, 3, 5, 10, 30, 500, and 20000, with 100 samplers targeting each of these temperatures. For a more complete discussion of parallel tempering in the current context, see Ref.~\onlinecite{webber2021rayleigh}. The temperatures are set to maintain the swapping rate between neighboring temperatures between $30\% \sim 40\%~$\cite{Predescu2005} (but the swapping rate between the last two ones are above $50\%$).
Our conclusions are robust to the choice and number of temperatures used.

\section{Umbrella sampling} \label{app:EMUS}

The eigenvector method for umbrella sampling (EMUS, \cite{dinner2020stratification}) is an enhanced sampling approach using a sequence of biasing functions $U_1, \ldots, U_L$ to produce low variance estimates of averages with respect to a given probability density $\pi(\boldsymbol{\sigma})$, i.e., estimate $\pi[g] := \Sigma_{\{\sigma_i\}} g(\boldsymbol{\sigma}) \pi(\boldsymbol{\sigma})  $. 
The steps of EMUS are as follows: 
\begin{algorithm}[EMUS]
~
\begin{enumerate}
    \item Sample from the distribution
    \begin{equation}
    \pi_i (\boldsymbol{\sigma})\propto \pi (\boldsymbol{\sigma})U_i(\boldsymbol{\sigma}),
    \end{equation}
    for $i = 1, \ldots, L$.
    \item Initialize 
    $\boldsymbol{u} = \begin{pmatrix} \frac{1}{L} & \cdots & \frac{1}{L} \end{pmatrix}$ 
    and repeat the following steps until the vector $\boldsymbol{u}$ converges:
    \begin{enumerate}
        \item Form the matrix $\mathbf{F}=\mathbf{F}(\boldsymbol{u})$ via 
        \begin{equation}
        F_{ij}=\pi_i \left[{\frac{U_j (\boldsymbol{\sigma})/u_j}{\sum_{k=1}^L U_k (\boldsymbol{\sigma})/u_k}}\right].
        \end{equation}
        \item Solve the eigenvalue problem
        \begin{equation}
        \boldsymbol{w}^T = \boldsymbol{w}^T \mathbf{F}(\boldsymbol{u}), \quad \sum_{i=1}^L w_i = 1.
        \end{equation}
        \item Set 
        \begin{equation}
        u_i = \frac{w_i u_i}{\sum_{j=1}^L w_j u_j}
        \end{equation}
        for $i = 1, \ldots, L$.
    \end{enumerate}
    \item Compute averages $\pi[g]$ as
    \begin{equation}\label{eq:usave}
    \pi[g]=\dfrac{\sum\limits_{i=1}^L w_i \pi_i \left[{\dfrac{g(\boldsymbol{\sigma})}{\sum U_k (\boldsymbol{\sigma})/u_k}}\right]}{ \sum\limits_{i=1}^L
    w_i \pi_i \left[{\dfrac{1}{\sum U_k (\boldsymbol{\sigma})/u_k}}\right]}.
    \end{equation}
\end{enumerate}
\end{algorithm}
We have used the symbol $\pi_i[\cdot]$ to denote averages with respect to the biased distributions $\pi_i$.
In this paper we apply the EMUS algorithm both (1) as one of the enhanced sampling methods of Sec.~\ref{sec:enhanced} and (2) as a tool for analyzing the VMC wave function, specifically the marginal density $P(s)$.

We now discuss the details of applications (1) and (2).
\begin{enumerate}
    \item \emph{Enhancing sampling.} In Sec.~\ref{sec:enhanced} we implement  umbrella sampling  with target distribution $\pi(\boldsymbol{\sigma}) \propto |\psi_\bt(\boldsymbol{\sigma})|^2$. 
We introduce $L = 16$ Gaussian biasing functions, defined by 
 \begin{equation}
\label{eq:gaussian}
    U_i(\boldsymbol{\sigma})
    = \exp\Bigl(-\frac{1}{2} \Bigl(\frac{s(\boldsymbol{\sigma}) - m_i}{\kappa}\Bigr)^2 \Bigr)
\end{equation}
for $i=2,...,15$,
\begin{equation*}
U_1(\boldsymbol{\sigma})
    = \begin{cases}
  1, & s(\bs)\leq m_1 \\
  \exp\Bigl(-\frac{1}{2} \Bigl(\frac{s(\boldsymbol{\sigma}) - m_1}{\kappa}\Bigr)^2 \Bigr), & s(\bs)> m_1
\end{cases}
\end{equation*}
and
\begin{equation*}
        U_{16}(\boldsymbol{\sigma})
    = \begin{cases}
  \exp\Bigl(-\frac{1}{2} \Bigl(\frac{s(\boldsymbol{\sigma}) - m_{16}}{\kappa}\Bigr)^2 \Bigr), & s(\bs)< m_{16}\\
    1,  & s(\bs)\geq m_{16} 
\end{cases}
\end{equation*}
with width parameter $\kappa = 0.1$ and centers $m_i$ given by 
$-0.5$, $-0.35$, $-0.15$, 0.05, 0.25, 0.45, 0.65, 0.85, 1.05, 1.3, 1.6, 1.9, 2.3, 2.7, 3.1, and 3.5 for $i=1,\ldots,16$, respectively. 
The centers are chosen to spread the MCMC samples across a broad range of $s$ values, while ensuring overlap between the MCMC samples in neighboring windows. Replica exchange moves between samples in neighboring windows are proposed every $2\times 10^{3}$ Metropolis steps~\citep{Matthews2018us}.
The swapping rates for these moves are $30\% \sim 40\%$, indicating sufficient overlap between windows.

    \item \emph{Computing the marginal density $P(s)$.} Taking 
    \begin{align}
        g_i(\bs)=|\psi_\bt(\bs)|^2\mathbbm{1}_{s(\bs)\in I_i}
    \end{align}
    for an interval of CV values $I_i$
    and $\pi$ to be the uniform distribution on spin configurations $\boldsymbol{\sigma}$ with an equal number of $+1$ and $-1$ spins, we use EMUS to estimate 
\begin{equation}
\label{eq:integrals}
    \pi[g_i] = \sum_{\{\sigma_j\}}^{s(\bs)\in I_i}
    |\psi_\bt(\boldsymbol{\sigma})|^2 
\end{equation}
 The CV marginal probability densities plotted in Fig.~\ref{fig:change_hist} in Sec.~\ref{sec:spurious}, Fig.~\ref{fig:CVUS_2} in Sec.~\ref{sec:enhanced}, and Fig.~\ref{fig:CV-VMC_VMC},~\ref{fig:CV-VMC_epsilon},~\ref{fig:CV-VMC_c} in Sec.~\ref{sec:numerical test} are estimated using this approach, where we define the intervals $I_i$ for $i=1,\ldots,17$ via 
 \begin{equation}
I_i = (-1.14 + 0.12 i, -1.02 + 0.12 i].
\end{equation}
For these experiments we use $L = 26$ Gaussian biasing functions of the same form as Eq.~\ref{eq:gaussian}
but with width $\kappa = 0.04$ and centers
\begin{equation}
    m_i = -1.08 + 0.08 i,
    \quad i = 1, \ldots, 26.
\end{equation}

\end{enumerate}
\vspace{2mm}

\section{CV-VMC sampling} \label{app:CV-VMCpool}

\begin{figure}[t!]
    \centering
    \includegraphics[width = 1.0\columnwidth]{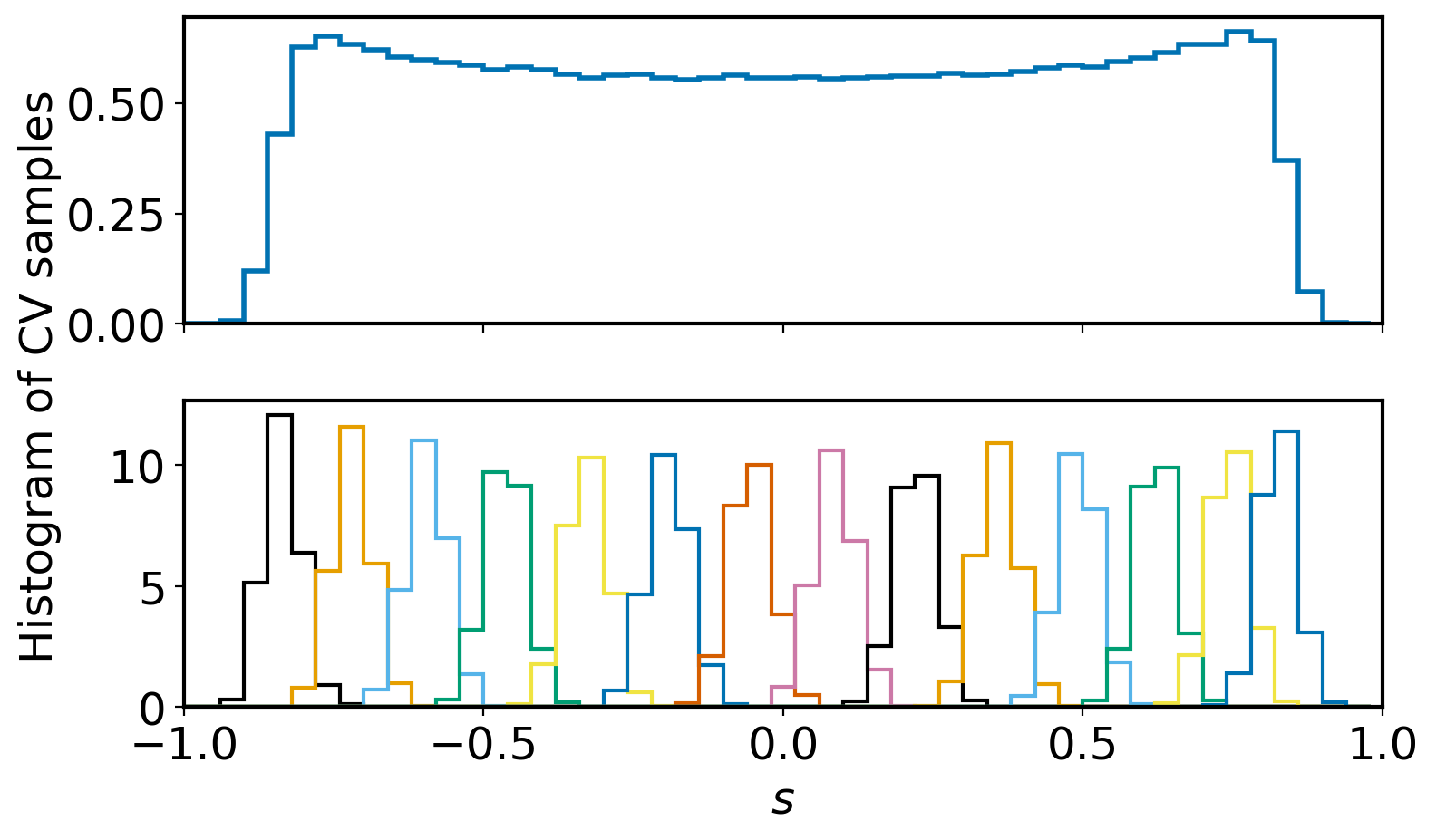}
    \caption{
    Top: histogram of samples used in the CV-VMC scheme.
    Bottom: histograms of MCMC samples in 14 representative windows (out of 52 total).}
    \label{fig:sumV}
\end{figure}

We use Gaussian biasing functions of the same form as Eq.~\ref{eq:gaussian} when preparing 
the pool of samples for the CV-VMC scheme in Sec.~\ref{sec:numerical test}.
To spread the samples to roughly evenly cover the range of possible CV  values, we use a greater number ($L = 52$) of Gaussian biasing functions, defined by width $\kappa = 0.04$ and centers
\begin{equation}
    m_i = -1.06 + 0.04 i,
    \qquad i = 1, \ldots, 52.
\end{equation}
These closely spaced biasing functions ensure that MCMC samples are evenly distributed across all values of $s$, as verified in Figure \ref{fig:sumV}. The reference probability density is chosen to be 
\begin{align}
    p(\bs)=\frac{1}{\sum_{i,\bs^{\prime}} U_i(\bs^{\prime})}\sum_i U_i(\bs).
\end{align}
In each window, 10 samplers start from uniformly distributed configurations within the subspace. We run $2\times 10^6$ Metropolis steps and collect one sample every $2\times 10^3$ steps. In this way we build a pool consisting of $5.2\times 10^{5}$ samples from $p$.

\section{Robustness test initialization}
\label{app:robust}
When training is carried out using standard Metropolis sampling or parallel tempering, the samples generated are distributed (approximately) according to $\rho_\bt\propto|\psi_\bt|^2$. 
However, when umbrella sampling is used to estimate averages with respect to $\pi= \rho_\bt$,
Eq.~\eqref{eq:usave} in Appendix~\ref{app:EMUS} shows that a sample at state $\boldsymbol{\sigma}$ drawn from the restrained distribution $\pi_i$ weighted by 
\[
\frac{w_i}{\sum U_k(\bs)/u_k}.
\]
Because of these weights, the samples used in training must be resampled to generate an unweighted set of samples for robustness test initialization (Sec.~\ref{sec:importance}).
Specifically, 
we select 100 of the final states of the MCMC samplers with replacement with probabilities proportional to
$w_i/\sum_k (U_k(\bs)/u_k)$ where $\bs$ is a state from the $i$th umbrella sampling window. 

\end{appendix}

\bibliography{references}

\end{document}